\documentclass[a4paper,aps,prd,10pt,preprintnumbers,showpacs,superscriptaddress,nofootinbib,amsmath,amssymb,floatfix
,twocolumn
]{revtex4-1}

\usepackage{enumitem}
\usepackage{graphicx}
\usepackage[utf8]{inputenc}
\usepackage[T1]{fontenc}
\usepackage{cmap}
\usepackage{graphicx}
\usepackage{dcolumn}
\usepackage{bm}
\usepackage{bbm}
\usepackage{amsfonts}
\usepackage{amsmath}
\usepackage{enumitem}
\usepackage{lipsum}
\usepackage[bottom]{footmisc}
\usepackage{cmap}
\usepackage{subcaption}
\usepackage{braket}
\usepackage{braket}
\usepackage[table]{xcolor}
\usepackage{amsmath}
\usepackage{hyperref}
\usepackage{cleveref}
\usepackage{fmtcount}
\usepackage{dblfnote}
\usepackage{float}
\usepackage{amssymb}
\usepackage{natbib}
\usepackage{blindtext}

\begin{document}

\title{Thermal, topological, and scattering effects of an AdS charged black hole with an antisymmetric tensor background}

\author{H. Chen}
\email{haochen1249@yeah.net}
\affiliation{School of Physics and Electronic Science, Zunyi Normal University, Zunyi 563006,PR China}


\author{M. Y. Zhang}
\email{myzhang94@yeah.net}
\affiliation{College of Computer and Information Engineering, Guizhou University of Commerce, Guiyang, 550014, China}


\author{A. A. Araújo Filho}
\email{dilto@fisica.ufc.br}
\affiliation{Departamento de Física, Universidade Federal da Paraíba, Caixa Postal 5008, 58051-970, João Pessoa, Paraíba, Brazil}


\author{F. Hosseinifar}
\email{f.hoseinifar94@gmail.com}
\affiliation{Department   of   Physics,   University   of   Hradec  Kr\'{a}lov\'{e}, Rokitansk\'{e}ho   62, 500   03   Hradec   Kr\'{a}lov\'{e},   Czechia}


\author{H. Hassanabadi}
\email{hha1349@gmail.com}
\affiliation{Department   of   Physics,   University   of   Hradec   Kr\'{a}lov\'{e}, Rokitansk\'{e}ho   62, 500   03   Hradec   Kr\'{a}lov\'{e},   Czechia}
\begin{abstract}
In this study, we explore a spherically symmetric charged black hole with a cosmological constant under the influence of a Kalb--Ramond field background. We compute the photon sphere and shadow radii, validating our findings using observational data from the Event Horizon Telescope (EHT), with a particular emphasis on the shadow images of Sagittarius $A^{*}$. Furthermore, we investigate the \textit{greybody} factors, emission rate, and partial absorption cross section. It is shown that the Lorentz-violating parameter \(\bar{l}\) has an important effect on the absorption cross section. Our analysis also includes an examination of the topological charge, temperature-dependent topology, and generalized free energy. In particular, we regard the AdS charged black hole with an antisymmetric tensor background as a topological defect in the thermodynamic space, then the system  has the same topological classification  to the charged RN-AdS black hole.
\end{abstract}
\maketitle


\section{Introduction}\label{Sec1}

The concept of Lorentz symmetry breaking has garnered substantial interest in contemporary physics, emerging as a crucial area of investigation across multiple theoretical paradigms \cite{bluhm2005spontaneous,bluhm2021gravity,kostelecky2011matter,kkos,tasson2014}. One notable method for exploring this phenomenon involves the introduction of the antisymmetric tensor field, commonly referred to as the Kalb--Ramond (KR) field \cite{altschul2010lorentz}. This field, rooted in string theory \cite{kalb1974classical}, provides a compelling approach for probing the intricacies of Lorentz symmetry violation.

The physics of black holes (BH) within the context of Kalb--Ramond (KR) gravity has been extensively studied \cite{17,18,19,vagnozzi2023horizon,1}. When the KR field is coupled with gravity, it can trigger spontaneous Lorentz symmetry breaking \cite{altschul2010lorentz}. This was demonstrated by deriving an exact solution for a static, spherically symmetric BH configuration . Building on this foundational work, researchers have explored the dynamics of both massive and massless particles near KR black holes \cite{5}. Studies have also focused on the gravitational deflection of light and the shadows produced by rotating black holes \cite{7}. Additionally, significant attention has been directed towards the detection of gravitational waves and their spectral characteristics within the framework of Lorentz symmetry breaking \cite{2,8,9,10,11}.

Moreover, the influence of Lorentz symmetry violation on electrically charged black holes has also been explored within the framework of KR gravity \cite{3}. For a static, spherically symmetric spacetime, the metric is described by:
\[
\mathrm{d}s^2=-f(r)\mathrm{d}t^2+\frac{1}{f(r)}\mathrm{d}r^2+h(r) \mathrm{d}\Omega^2\label{ds2},
\]
where, in agreement with Ref. \cite{3}:
\begin{eqnarray}
f(r)=\frac{1}{1-\bar{l}}-\frac{2 M}{r}+\frac{Q}{(1-\bar{l})^2 r^2}-\frac{\Lambda  r^2}{3 (1-\bar{l})}.\label{fr}
\end{eqnarray}
Here, \( M \) represents the black hole mass, \( Q \) denotes the electric charge, \( \Lambda \) is the cosmological constant, and \(\bar{l}\) is the Lorentz--violating parameter, with \( h(r) = r^2 \). In this context, various phenomena such as the evaporation process, quasinormal modes, and emission rates have been studied \cite{2}. Additionally, numerous aspects of black hole physics have been investigated for the case of a zero electric charge, as discussed in Refs. \cite{13,14,15,16,araujo2024exploring}.

In particular, we investigate a spherically symmetric charged black hole with a cosmological constant in the presence of a KB field background. We calculate the photon sphere and shadow radii and validate our results using observational data from the Event Horizon Telescope (EHT), with a particular focus on the shadow images of Sagittarius \(A^{*}\). Additionally, we examine \textit{greybody} factors, emission rates, and partial absorption cross sections. Our analysis extends to exploring the topological charge, temperature-dependent topology, and generalized free energy.

This work is structured as follows: In Sec. \ref{Sec2}, we calculate the shadow radii and derive constraints based on observational data from Sagittarius \(A^{*}\). Sec. \ref{Sec3} focuses on the heat capacity of the system. In Sec. \ref{Sec4}, we discuss the \textit{greybody} bounds, highlighting relevant emission power and the absorption cross section. Sections \ref{Sec5}, and \ref{Sec6} explore the topological aspects of the black hole and its charge. Finally, in Sec. \ref{Sec9}, we summarize our findings and present concluding remarks.


\section{Shadow Radius}\label{Sec2}

The behavior of light can be determined by applying the Euler--Lagrange equation as follows:
\begin{eqnarray}
\frac{\mathrm{d}}{\mathrm{d}\tau}\left(\frac{\partial \mathcal{L}}{\partial \dot{x}^{\mu}}\right)-\frac{\partial \mathcal{L}}{\partial x^{\mu}}=0,\,\,\,\,\, \mathcal{L}=\frac{1}{2}g_{\mu\nu}\dot{x}^{\mu}\dot{x}^{\nu}.
\end{eqnarray}
At \(\theta = \pi/2\), we are able to define two constants of motion: \(L = h(r) \dot{\phi}\) and \(E = f(r) \dot{t}\) so that effective potential reads
\cite{chen2024quasi}
\begin{eqnarray}
V_{eff} = \frac{f(r)}{h(r)}\bigg(\frac{L^2}{E^2}-1\bigg)
\end{eqnarray}
is acquired. Given that a circular orbit requires \(V_{eff}(r) = 0\) and \(V'_{eff}(r) = 0\), the behavior of a photon, particularly the photon radius around a black hole, can be determined from these conditions \cite{29,30}
\begin{eqnarray}
r_{ph} \partial_r f(r)\bigg|_{r=r_{ph}}-2f(r_{ph})=0,
\end{eqnarray}
in this manner, the photon radius is given by
\begin{eqnarray}
r_{ph}=-\frac{\sqrt{9 (\bar{l}-1)^4 M^2+8 (\bar{l}-1) Q}+3 (\bar{l}-1)^2 M}{2 (\bar{l}-1)}.\label{rph}
\end{eqnarray}
Indeed, the observed shadow of a black hole appears distorted and larger than its actual size and shape \cite{Perlick}. Owing to the geometry, for an observer situated at a distance \(r_o\) from the black hole, the shadow radius is given by \cite{araujo2024dark}
\begin{align}
\nonumber
r_{sh}&= r_{ph}\frac{\sqrt{f(r_{o})}}{\sqrt{f(r_{r_{ph}})}}\\
&=\frac{r_{ph}^2 \sqrt{3 Q-(\bar{l}-1) r_o \left(6 (\bar{l}-1) M-\Lambda  r_o^3+3 r_o\right)}}{r_o \sqrt{3 Q-(\bar{l}-1) r_{ph} \left(6 (\bar{l}-1) M-\Lambda  r_{ph}^3+3 r_{ph}\right)}},
\end{align}

\begin{figure}[ht!]
\centering
	\begin{subfigure}[b]{0.5\textwidth}
	\centering
	\includegraphics[scale = 0.48]{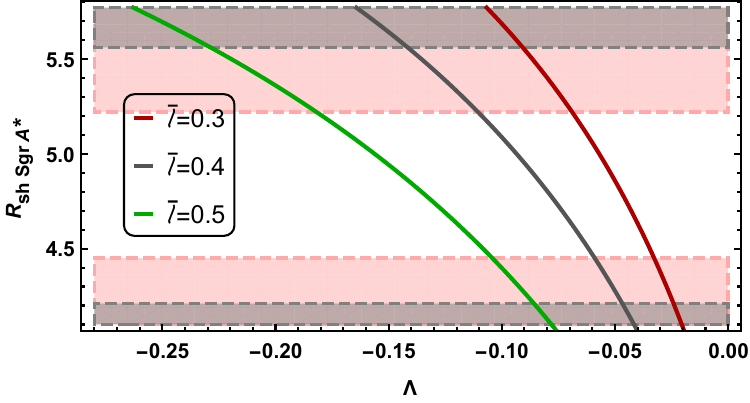} \hspace{-0.2cm}\\
	\caption{$Q=0.01$}
    \end{subfigure}%
\hfill
\centering
	\begin{subfigure}[b]{0.5\textwidth}
	\centering
	\includegraphics[scale = 0.48]{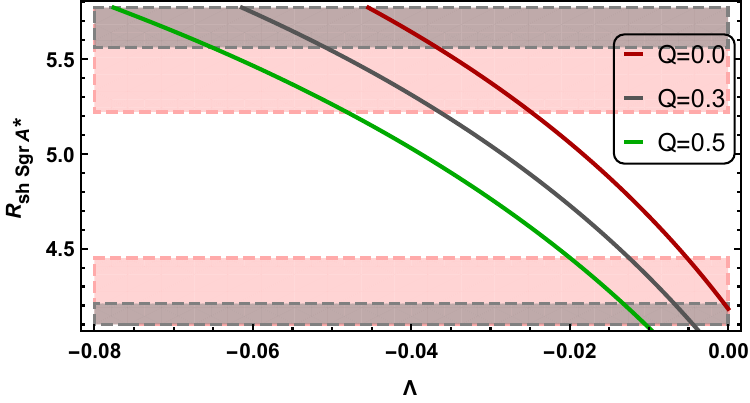} \hspace{-0.2cm}\\
	\caption{$\bar{l}=0.1$}
    \end{subfigure}%
\caption{The shadow radius as a function of the parameter \(\Lambda\) is illustrated. The white region indicates the permissible range for \(1\sigma\). Both the pink and white regions are acceptable for \(2\sigma\).}
\label{fig:shadowLambda}
\end{figure}

\begin{figure}[ht!]
\centering
	\begin{subfigure}[b]{0.5\textwidth}
	\centering
	\includegraphics[scale = 0.48]{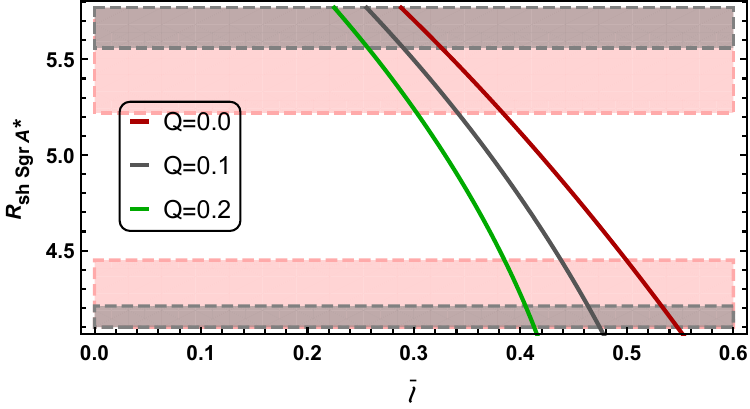} \hspace{-0.2cm}\\
	\caption{$\Lambda=-0.1$}
    \end{subfigure}%
\hfill
\centering
	\begin{subfigure}[b]{0.5\textwidth}
	\centering
	\includegraphics[scale = 0.48]{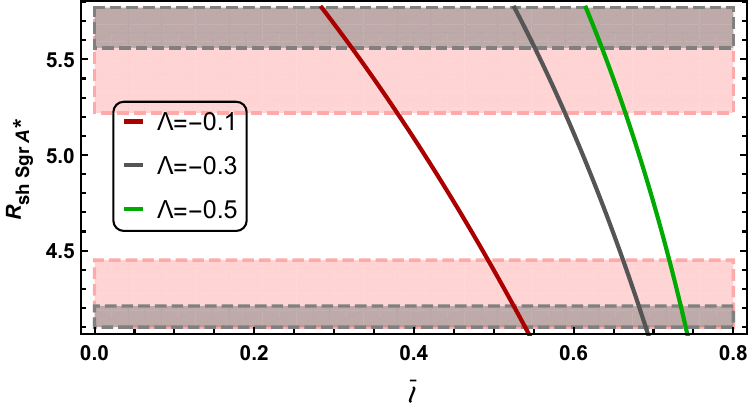} \hspace{-0.2cm}\\
	\caption{$Q=0.01$}
    \end{subfigure}%
\caption{The shadow radius as a function of the parameter \(l\) is shown. The pink area is excluded for \(1\sigma\), and the gray region is not acceptable for \(2\sigma\).}
\label{fig:shadowl}
\end{figure}

\begin{figure}[H]
\centering
	\begin{subfigure}[b]{0.5\textwidth}
	\centering
	\includegraphics[scale = 0.48]{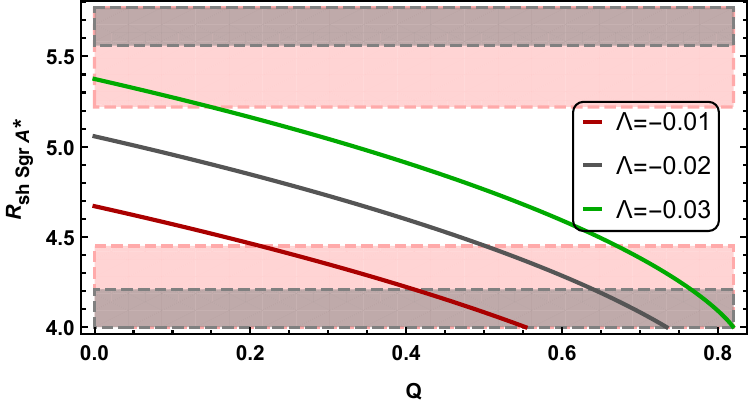} \hspace{-0.2cm}\\
	\caption{$\bar{l}=0.1$}
    \end{subfigure}%
\hfill
\centering
	\begin{subfigure}[b]{0.5\textwidth}
	\centering
	\includegraphics[scale = 0.48]{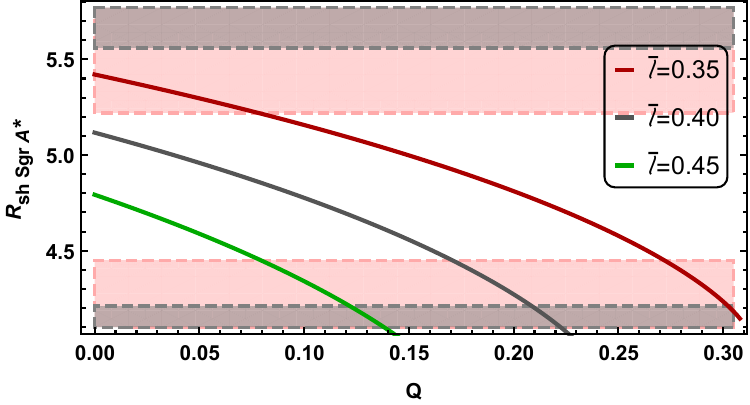} \hspace{-0.2cm}\\
	\caption{$\Lambda=-0.1$}
    \end{subfigure}%
\caption{The shadow radius as a function of the parameter \(Q\) is depicted. The pink area is excluded for \(1\sigma\), and the gray region is not acceptable for \(2\sigma\).}
\label{fig:shadowQ}
\end{figure}

According to the shadow radius curves shown in FIG \ref{fig:shadowLambda} to \ref{fig:shadowQ}, the permissible values for the parameters, ensuring the shadow radius remains within the acceptable range, are listed in Table \ref{Table:rsh}.

\begin{figure*}
\centering
\begin{table}[H]
		\caption{The upper and lower bounds of the parameters are determined based on observations of Sgr A$^*$.}
		\centering
		\label{Table:rsh}
		\begin{tabular*}{\textwidth}{@{\extracolsep{\fill}}|ccccc|ccccc|}
		\hline
		\hline
		 & \multicolumn{2}{c}{$1\sigma$} & \multicolumn{2}{c|}{$2\sigma$} & & \multicolumn{2}{c}{$1\sigma$} & \multicolumn{2}{c|}{$2\sigma$}\\
		 \hline
		 $\Lambda\,(Q=0.01)$ & Lower & Upper & Lower & Upper & $\Lambda\,(\bar{l}=0.10)$ & Lower & Upper & Lower & Upper\\
		  \hline
		 $\bar{l}=0.30$ & $-0.069$ & $-0.037$ & $-0.091$ & $-0.024$ & $Q=0.0$ & $-0.025$ & $-0.007$ & $-0.037$ & $-0.001$\\
		 \hline
		 $\bar{l}=0.40$ & $-0.111$ & $-0.065$ & $-0.142$ & $-0.047$ & $Q=0.3$ & $-0.037$ & $-0.015$ & $-0.051$ & $-0.007$\\
		 \hline
		 $\bar{l}=0.50$ & $-0.182$ & $-0.113$ & $-0.229$ & $-0.086$ & $Q=0.5$ & $-0.048$ & $-0.023$ & $-0.065$ & $-0.013$\\
		 \hline
		 \hline
		 $\bar{l}\,(\Lambda=-0.10)$ & Lower& Upper& Lower & Upper & $\bar{l}\,(Q=0.01)$ & Lower& Upper& Lower & Upper\\
		  \hline
		 $Q=0.00$ & $0.383$ & $0.485$ & $0.326$ & $0.533$ & $\Lambda=-0.10$ & $0.379$ & $0.479$ & $0.322$ & $0.525$\\
		 \hline
		 $Q=0.10$ & $0.341$ & $0.427$ & $0.326$ & $0.533$ & $\Lambda=-0.30$ & $0.589$ & $0.653$ & $0.551$ & $0.682$\\
		 \hline
		 $Q=0.20$ & $0.303$ & $0.376$ & $0.257$ & $0.405$ & $\Lambda=-0.50$ & $0.665$ & $0.713$ & $0.635$ & $0.734$\\
		 \hline
		 \hline
		 $Q\,(\bar{l}=0.10)$ & Lower& Upper& Lower & Upper & $Q\,(\Lambda=-0.10)$ & Lower& Upper& Lower & Upper\\
		  \hline
		 $\Lambda=-0.01$ & $--$ & $0.119$ & $--$ & $0.409$ & $\bar{l}=0.35$ & $0.078$ & $0.255$ & $--$ & $0.302$\\
		 \hline
		 $\Lambda=-0.02$ & $--$ & $0.434$ & $--$ & $0.640$ & $\bar{l}=0.40$ & $--$ & $0.151$ & $--$ & $0.209$\\
		 \hline
		 $\Lambda=-0.03$ & $0.149$ & $0.622$ & $--$ & $0.684$ & $\bar{l}=0.45$ & $--$ & $0.058$ & $--$ & $0.122$\\
		 \hline
		 \hline
		\end{tabular*}
	\end{table}
	\end{figure*}

Notice that an increase in the parameter \(\bar{l}\) results in a decrease in the value of the shadow radius.


\section{Heat Capacity}\label{Sec3}

\begin{figure}[ht!]
\centering
	\begin{subfigure}[b]{0.5\textwidth}
	\centering
	\includegraphics[scale = 0.48]{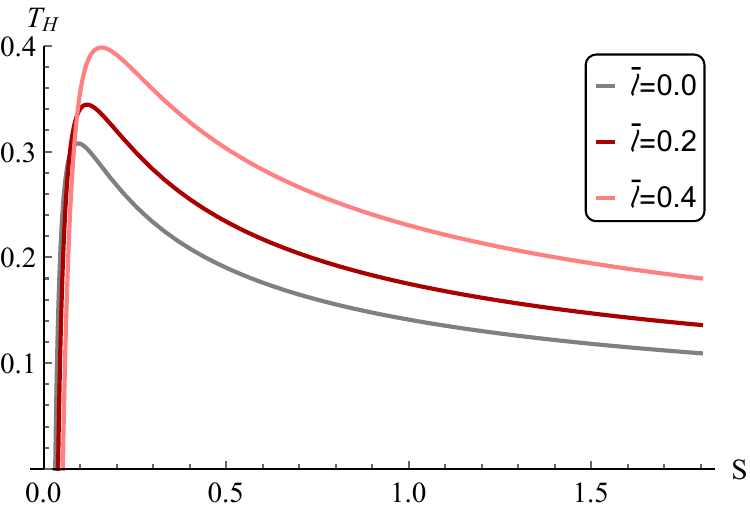} \hspace{-0.2cm}\\
    \end{subfigure}%
    \caption{The relationship between temperature and entropy is depicted, with the parameters set to \(Q = 0.01\) and \(\Lambda = -0.1\).}\label{fig:TS}
\end{figure}

As it is well know, the \textit{Hawking} temperature of a  corresponding black hole is
\begin{eqnarray}
T_H=\partial_r f(r)\big|_{r=rh},
\end{eqnarray}
where \(r_h\) refers to the horizon radius so that, for our case under consideration, it follows \cite{3}
\begin{eqnarray}
T_H=\frac{ r_h^2 (\bar{l}-1) \left(\Lambda  r_h^2-1\right)-Q}{4 \pi  (\bar{l}-1)^2 r_h^3}.\label{TH}
\end{eqnarray}
In addition, the entropy of the system is \cite{31}
\begin{eqnarray}
S=\pi r_h^2.
\end{eqnarray}
FIG. \ref{fig:TS} illustrates the \(T-S\) curve. It is evident that as the parameter \(\bar{l}\) increases, the maximum value of temperature, along with its corresponding entropy, shifts to higher values.

For the sake of completeness of our study, we also present the heat capacity
\begin{align}
\nonumber
C=& T_H \frac{\partial S}{\partial T_H}
\\=&\frac{2 \pi  r_h^2 \left((\bar{l}-1) r_h^2 \left(\Lambda  r_h^2-1\right)-Q\right)}{(\bar{l}-1) r_h^2 \left(\Lambda  r_h^2+1\right)+3 Q}.
\end{align}

\begin{figure}[ht!]
\centering
	\begin{subfigure}[b]{0.5\textwidth}
	\centering
	\includegraphics[scale = 0.48]{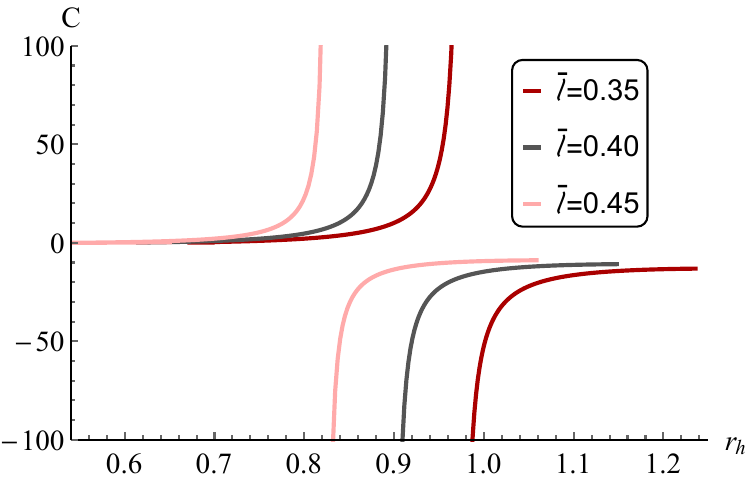} \hspace{-0.2cm}\\
    \end{subfigure}
\caption{The relation between heat capacity and \(r_h\) is presented, where \(r_h\) is calculated by varying the parameter \(Q\). The parameters are set to \(\Lambda = -0.1\).
}
\label{fig:Cv}
\end{figure}

FIG. \ref{fig:Cv} illustrates the behavior of heat capacity as a function of horizon radius. The curve indicates that as the parameter \(\bar{l}\) increases, the phase transition occurs at a larger horizon radius, leading to a negative heat capacity. It is worth mentioning that such negative values signifies instability, whereas a positive heat capacity indicates a thermodynamically stable black hole.

\section{Greybody}\label{Sec4}

\begin{figure}[ht!]
\centering
	\begin{subfigure}[b]{0.5\textwidth}
	\centering
	\includegraphics[scale = 0.48]{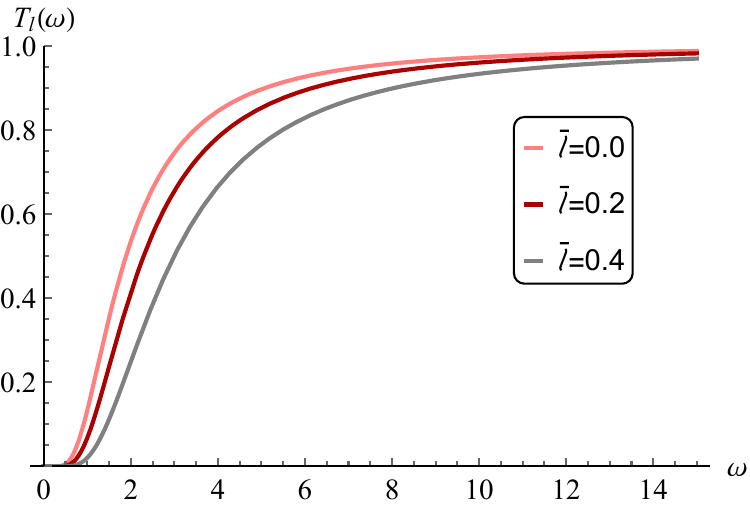} \hspace{-0.2cm}
    \end{subfigure}%
\hfill
\caption{The \textit{greybody} bounds for different values of \(l\) are presented, with \(l = 2\) and the other parameters set as \(M = 1\), \(\Lambda = -0.1\), \(Q = 0.01\), and \(s = 1\).}
\label{fig:gbb}
\end{figure}

The \textit{greybody} effect refers to the radiation emitted as a result of the \textit{Hawking} temperature. The \textit{greybody} bound signifies the maximum deviation of this radiation from the ideal blackbody spectrum and is calculated from \cite{konoplya2023bardeen,110,11,112,heidari2024scattering}
\begin{eqnarray}
T_l(\omega)\geq \text{sech}^2 \left(\frac{1}{2\omega}\int_{r_h}^{\infty}V(r) \frac{dr}{f(r)}\right)\label{gbb},
\end{eqnarray}
where the potential is
\begin{eqnarray}
V(r)=f(r) \left(\frac{(1-s^2)\partial_r f(r)}{r}+\frac{(l+1) l}{r^2}\right)\label{potential}.
\end{eqnarray}
Here, \(l\) represents the angular momentum, while \(s\) refers to the multipole number. Specifically, \(s = 0\) is associated with scalar perturbations, and \(s = 1\) corresponds to electromagnetic perturbations. The \textit{greybody} bounds for several initial values and three different \(\bar{l}\) parameters are illustrated in FIG. \ref{fig:gbb}. It is evident that as the parameter \(\bar{l}\) increases, the probability curve shifts upward.
The $l$th mode emitted power is given by \cite{16}
\begin{eqnarray}
P_{l}(\omega)=\frac{A}{8\pi^2}T_l(\omega)\frac{\omega^3}{\exp(\omega / T_H)-1},
\end{eqnarray}
where $A$ represents the surface area.
\begin{figure}[ht!]
\centering
	\begin{subfigure}[b]{0.5\textwidth}
	\centering
	\includegraphics[scale = 0.48]{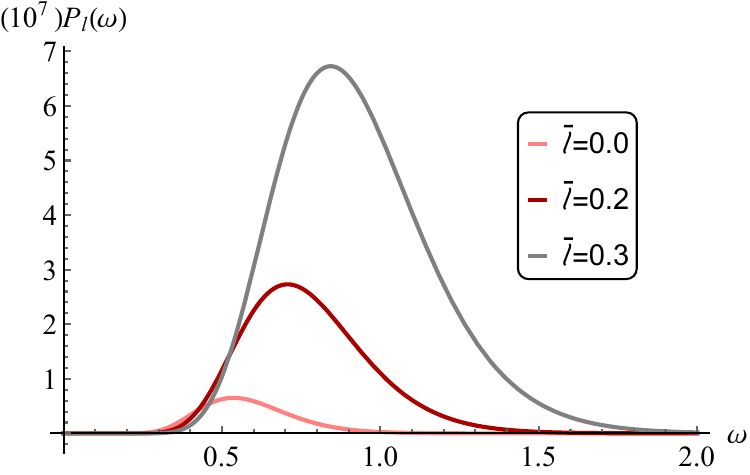} \hspace{-0.2cm}\\
    \end{subfigure}%
\hfill
\caption{The emitted power for three values of $\hat{l}$ is illustrated, with $l = 2$ and the other parameters set as \(M = 1\), \(\Lambda = -0.1\), \(Q = 0.01\), and \(s = 1\)}.
\label{fig:gbbpower}
\end{figure}
FIG. \ref{fig:gbbpower} illustrates the emission power, clearly showing that as the parameter \(\bar{l}\) increases, both the maximum value and its corresponding frequency decrease.

\begin{figure}[ht!]
\centering
	\begin{subfigure}[b]{0.5\textwidth}
	\centering
	\includegraphics[scale = 0.48]{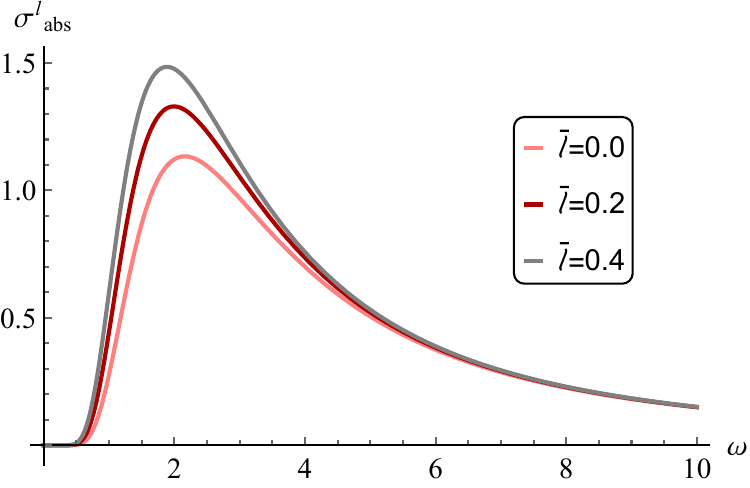} \hspace{-0.2cm}
    \end{subfigure}%
\hfill
\caption{The absorption cross--section curve for different values of \(\hat{l}\) is presented, with \(l = 2\) and the other parameters set as \(M = 1\), \(\Lambda = -0.1\), \(Q = 0.01\), and \(s = 1\).}
\label{fig:gbb_sigma}
\end{figure}

Furthermore, the absorption cross section written below \cite{27,28}
\begin{eqnarray}
\sigma_{abs}^l =\frac{\pi (2l+1)}{\omega^2} |T_l(\omega)|^2.
\end{eqnarray}
The absorption cross--section is depicted in FIG. \ref{fig:gbb_sigma}. As observed, varying the parameter \(\bar{l}\) results in a shift of the curve. An increase in this parameter leads to a higher maximum value for absorption, which occurs at a lower frequency.


\section{Topological Charge of Photon Sphere}\label{Sec5}
Recently, Wei al. creatively applied the  topological current theory to black hole thermodynamics \cite{0028}. In this direction, the thermodynamic topology of black holes in different backgrounds is investigated \cite{l0,l1,l2,l3,l4,l5,l7,l8,l9,l10,l11,l12,l13,l14,l15,l16,l17,l18,l19,l20,l21,l22,l23,l24,l25,l26,l27,l28}. We will explore the topological characteristics of the photon sphere. We begin by defining the everywhere regular potential function as \cite{23,24}
\begin{eqnarray}
H(r,\theta)=\frac{1}{\sin\theta}\sqrt{\frac{f(r)}{h(r)}},
\end{eqnarray}
Here, we assume \( h(r) = r^2 \). The root of \(\partial_r H = 0\) corresponds to the radius of the photon sphere. To determine the topological charge associated with it, we define a vector field \(\phi = (\phi_{r}, \phi_{\theta})\), as described in \cite{23}.
\refstepcounter{equation}
\begin{align}
\phi_{r}=\sqrt{f(r)} \partial_r H(r,\theta)\tag{\theequation a},\\
\phi_{\theta}=\frac{1}{\sqrt{h(r)}}\partial_{\theta} H(r,\theta)\tag{\theequation a}.
\end{align}
The vector $\phi$ can be also rewrite as $\phi=||\phi||e^{i\Theta}$, where $||\phi||=\sqrt{\phi_a\phi_a} $, $a$=1, 2, and $\phi_1=\phi_r$, $\phi_2=\phi_\theta$. It is important to note that the zero point of \(\phi\) coincides precisely with the location of the photon sphere. This implies that \(\phi\) in \(\phi = ||\phi|| e^{i\Theta}\) is not well-defined at this point. Therefore, the vector is considered as \(\phi = \phi_r + i \phi_\Theta\). The normalized vectors are defined as follows:
\begin{eqnarray}
n_r=\frac{\phi_{r}}{||\phi||},\;n_{\theta}=\frac{\phi_{\theta}}{||\phi||}.
\end{eqnarray}
Moreover, the topological current is established as follows
\begin{equation}\label{T2}
J^\mu=\frac{1}{2\pi } \epsilon ^{\mu \nu \lambda }\epsilon _{ab}\partial_{ \nu }n_a\partial_{ \lambda  }n_b,
\end{equation}
Since \(\epsilon^{\mu \nu \lambda} = -\epsilon^{\mu \lambda \nu}\), it can be easily verified that the topological current is conserved, satisfying \(\partial_\mu J^\mu = 0\). The zero component of the topological current is denoted by \(J^0\), and integrating this component over a specified region yields the total topological charge.
\begin{equation}\label{T5}
\begin{aligned}
\mathfrak{Q} =\int_{\sum }^{} j^0d^2x=\sum_{i=1}^{N}  \beta_i\eta_i=\sum_{i=1}^{N}  w_i,
\end{aligned}
\end{equation}
where, \(\beta_i\) and \(\eta_i\) represent the Hopf index and Brouwer degree at the zero point \(z_n\), respectively. Considering a closed, smooth, and positively oriented loop \(C_i\) that encircles the \(i\)-th zero point of \(\phi\) while excluding other zero points, the winding number of the vector is given by
\begin{equation}
\begin{aligned}
w_i  & =\frac{1}{2 \pi} \oint_{C_i} d \Omega ,
\end{aligned}
\end{equation}
due to the geometry, we can obtain
\begin{eqnarray}
\Omega=\arctan \frac{\phi_\theta}{\phi_r}=\arctan\frac{n_\theta}{n_r}.
\end{eqnarray}
In this manner,
\begin{eqnarray}
d \Omega=\frac{n_r \partial_{\theta} n_{\theta}-n_{\theta} \partial_{\theta} n_r}{n_{r}^2+n_{\theta}^{2}},
\end{eqnarray}
and the topological charge can be calculated from
\begin{eqnarray}
\mathfrak{Q}=\frac{\Delta\Omega}{2\pi}.
\end{eqnarray}

\begin{figure}[ht!]
\centering
	\begin{subfigure}[b]{0.5\textwidth}
	\centering
	\includegraphics[scale = 0.48]{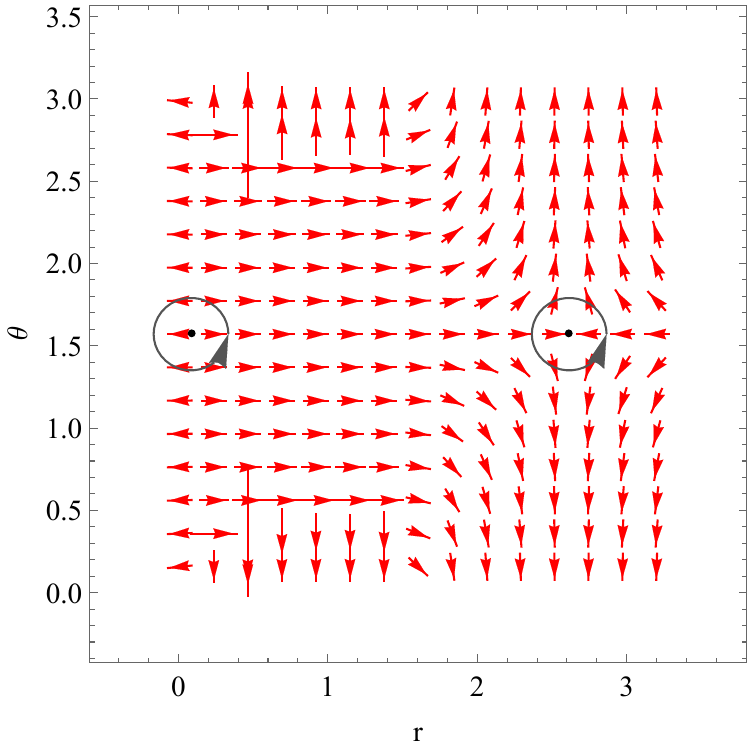} \hspace{-0.2cm}\\
	\caption{Parameters are set to $\Lambda =-0.1,\,Q=0.1,\,\bar{l}=0.1$. At $r = 2.61502,\, w=-1$ and at $r = 0.0849791,\, w=1$}
	\hspace{-0.2cm}\\
    \end{subfigure}%
    \hfill
	\begin{subfigure}[b]{0.5\textwidth}
	\centering
	\includegraphics[scale = 0.48]{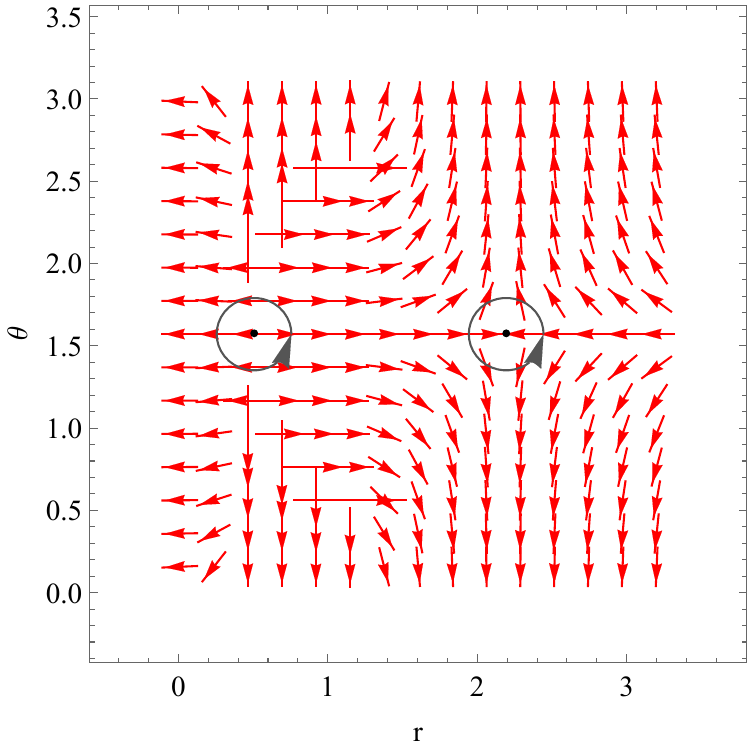} \hspace{-0.2cm}\\
	\caption{Parameters are set to $\Lambda =-0.1,\,Q=0.5,\,l=0.1$. At $r = 2.19344$ $w=-1$ and at $r = 0.506561$, $w=1$.}
	\hspace{-0.2cm}\\
    \end{subfigure}%
    \hfill
\caption{The topological charge of photon spheres.}
\label{fig:phasetr}
\end{figure}

The poles of \(\Delta\Omega\) occur at the photon radius, and \(\mathfrak{Q}\) can take values of \(0, \pm 1\). The illustration of the vector space \((n_r, n_{\theta})\) is shown in Fig. \ref{fig:phasetr}. In FIG. \ref{fig:phasetr}(a), there is a photon sphere located at \((r, \theta) = (2.61502, \pi/2)\), around which the field lines converge towards the zero point of the vector field, resembling the electric field generated by a negative charge and possessing a topological charge of $-1$. Based on the classification in \cite{23}, this photon sphere is considered standard and unstable.

An exotic photon sphere exists at \((r, \theta) = (0.0849791, \pi/2)\), where the field lines diverge near the zero point, similar to the electric field produced by a positive charge, with a topological charge of +1. This photon sphere is stable and corresponds to the region of the naked singularity \cite{0028}. As the charge \(Q\) increases, as depicted in FIG. \ref{fig:phasetr}(b), the standard photon sphere and exotic photon sphere approach one another. When \(Q\) increases to approximately $0.82$, the two photon spheres converge. Further increases in \(Q\) lead to the absence of rings in spacetime, resulting in a total topological charge of \(\mathfrak{Q} = 0\).


\section{Topology in Temperature and  Generalized Free Energy}\label{Sec6}
\begin{figure}[ht!]
\centering
	\begin{subfigure}[b]{0.5\textwidth}
	\centering
	\includegraphics[scale = 0.35]{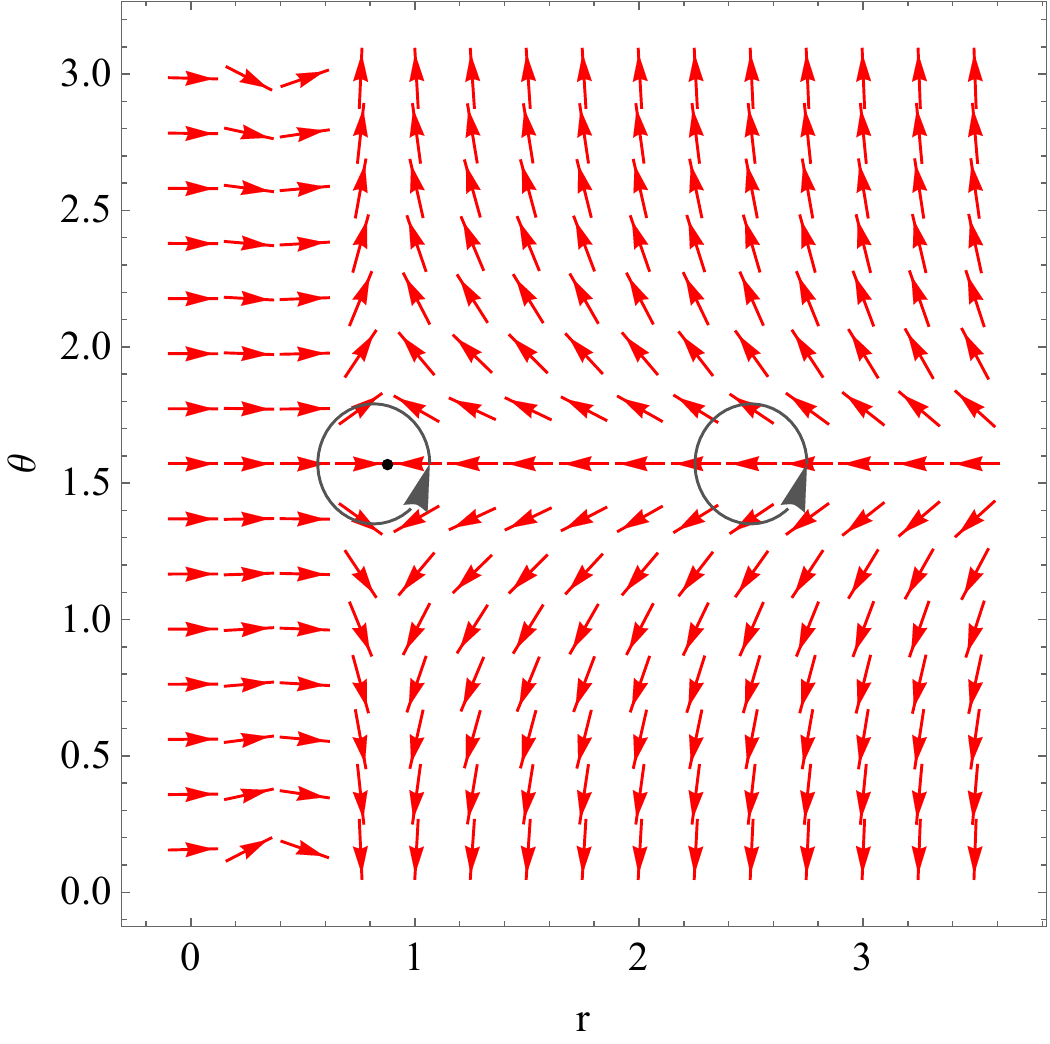} \hspace{-0.2cm}\\
	\caption{Isobaric curves and spinodal curve (gray line).}
    \end{subfigure}%
\hfill
	\begin{subfigure}[b]{0.5\textwidth}
	\centering
	\includegraphics[scale = 0.5]{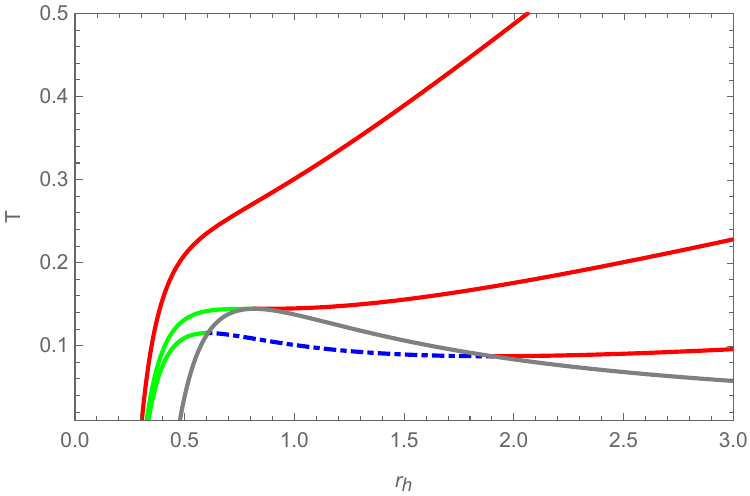} \hspace{-0.2cm}\\
	\caption{The topological charge for $Q=0.1$ and $\bar{l}=0.1$ at $r=0.8165$, $w=-1$.}
    \end{subfigure}%
\hfill
\caption{Temperature topological charge for $Q=0.1$ and $\bar{l}=0.1$ at $r=0.8165$.}
\label{fig:PhaseTemp}
\end{figure}

\begin{figure}
\centering
	\begin{subfigure}[b]{0.5\textwidth}
	\centering
	\includegraphics[scale = 0.5]{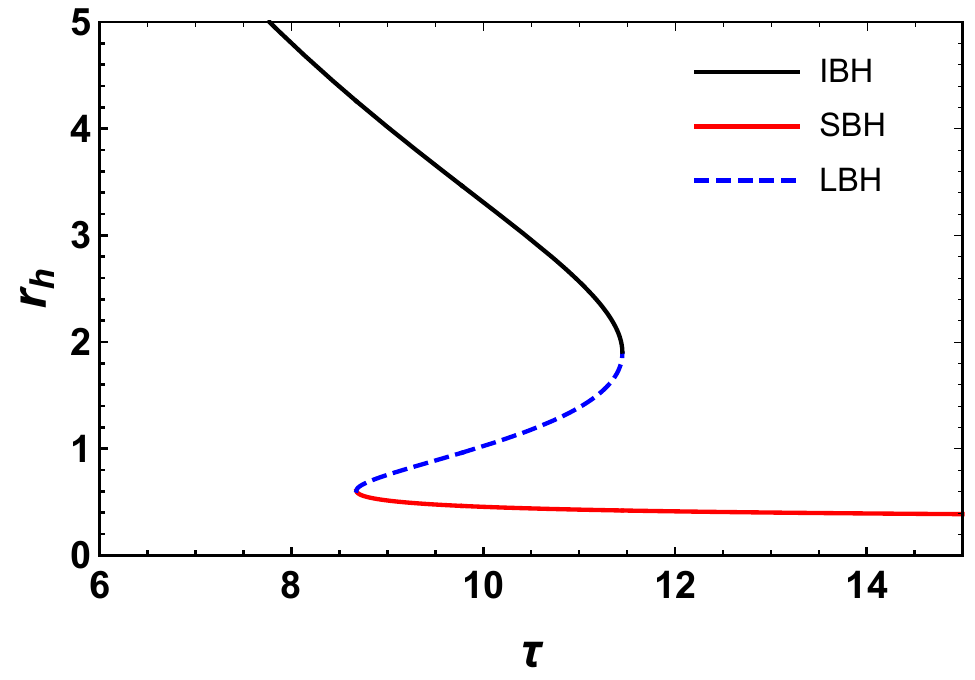} \hspace{-0.2cm}\\
	\caption{The zero points of the vector field $\phi$ for $Q=0.1$, $\bar{l}=0.1$ and $P=0.01$ ($P<P_c$). Critical points are $(8.67847,\,0.60599)$, and $(11.4513,\,1.90043)$.}
    \end{subfigure}%
\hfill
	\begin{subfigure}[b]{0.5\textwidth}
	\centering
	\includegraphics[scale = 0.42]{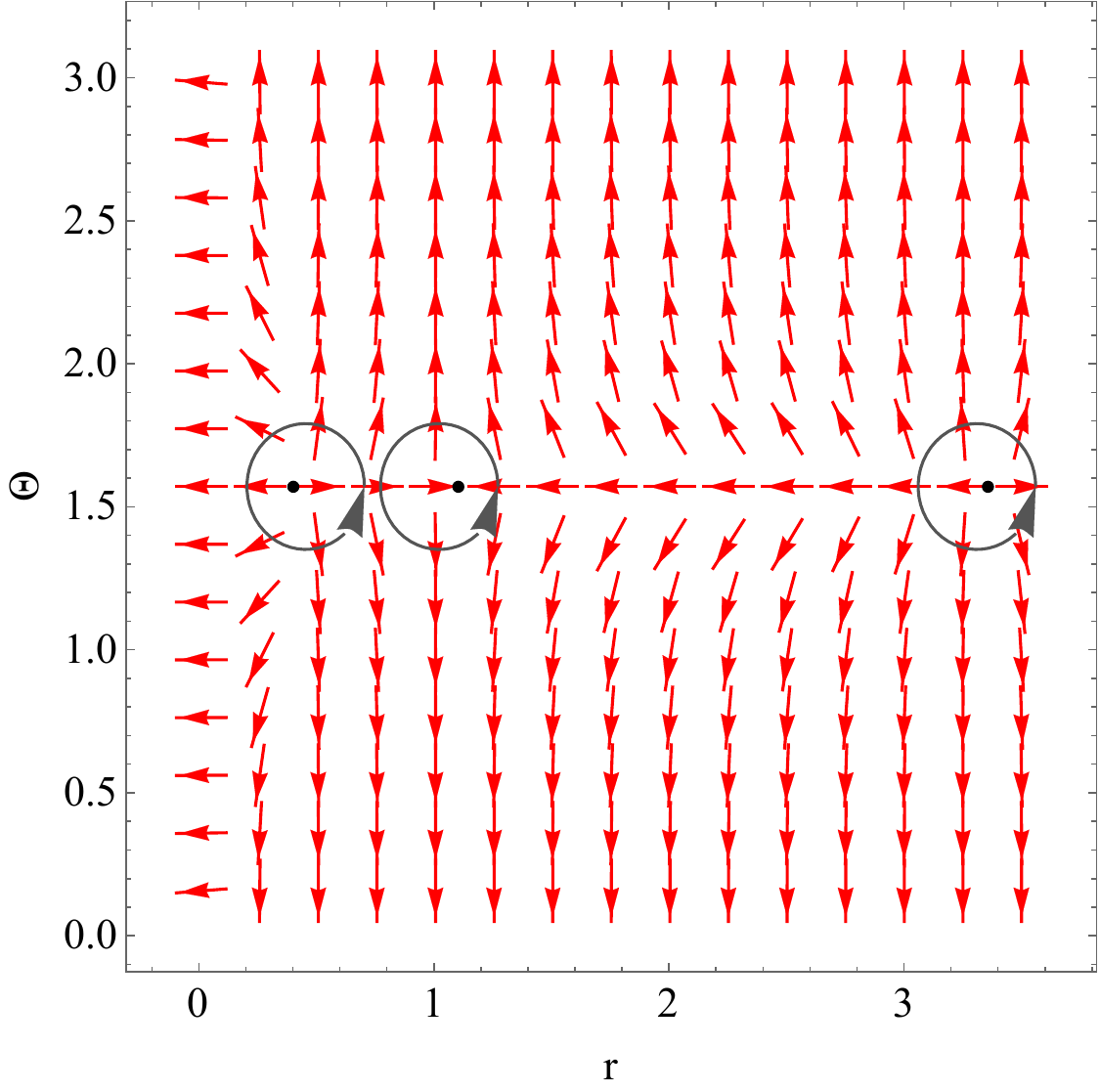} \hspace{-0.2cm}\\
	\caption{The topological numbers for $Q=0.1$, $\bar{l}=0.1$, $P=0.01$ and $\tau=10$. $r=0.454497$, $w=1$, $r=1.02267$, $w=-1$ and $r=3.31018$, $w=1$.}
    \end{subfigure}%
\hfill
\caption{The topological numbers of the black holes solutions.}
\label{fig:PhaseF}
\end{figure}

In this section, we will apply temperature and generalized free energy methods to analyze the topological structure of charged spherically symmetric black holes. The critical pressure, given by \(\Lambda = -8 \pi P\), can be found by setting \(\partial_{r_h} T_H \big|_{P=P_c} = 0\). By substituting \(P_c\) into the \textit{Hawking} temperature in the form given by equation \ref{TH}, we can rewrite \(T_H\) as \(\tilde{T}_H\). Consequently, a field can be defined as follows \cite{0028}:
\begin{align}\nonumber
\Phi =&\frac{1}{\sin (\theta)} \tilde{T}_H\\=&-\frac{\csc (\theta ) \left((\bar{l}-1) r_h^2 \left(\dfrac{\bar{l} r_h^2+3 Q-r_h^2}{(\bar{l}-1) r_h^2}+1\right)+Q\right)}{4 \pi  (\bar{l}-1)^2 r_h^3}.
\end{align}
Since vector fields on a two-dimensional plane are more intuitive than those in one or higher dimensions, \(\theta\) serves as an auxiliary factor that aids in topological analysis. The unit vectors of this field read
\refstepcounter{equation}
\begin{align}
\varphi_r&=\partial_{r_h}\Phi\tag{\theequation a},\\\varphi_{\theta}&=\partial_{\theta}\Phi\tag{\theequation b}.
\end{align}
Specifically, when \(\theta = \pi/2\), the vector \(\phi = (\phi_r, \phi_{\theta})\) is always zero. It is straightforward to see that the critical point coincides with the zero point of \(\phi\). The normalized vector \(n^a = \frac{\phi^a}{\|\phi\|}, \, (a=1,2)\) is plotted in FIG. \ref{fig:PhaseTemp}(a), which illustrates only one critical point. Two contours will be constructed: one enclosing the critical point, which has a topological charge of $-1$, and another that does not enclose any critical points, possessing a topological charge of $0$. At this stage, the critical point is conventional and corresponds to the maximum of the spinodal curve in the isobaric diagram, as shown in the figure.
\ref{fig:PhaseTemp}(b).
When the charged spherically symmetric black hole solution acts as a defect in the thermodynamic parameter space, the generalized Helmholtz free energy is expressed as follows: \cite{0029}
\begin{align}\nonumber
F=&M(r_h)-\frac{S} {\tau }\\=&\frac{3Q+3r_h^2-3\bar{l} r_h^2-r_h^4\Lambda +\bar{l} r_h^4\Lambda}{6(\bar{l}-1)^2r_h}-\frac{\pi r_h^2} {\tau },
\end{align}
where \(\Lambda = -8\pi P\). This free energy exhibits its on-shell characteristics when \(\tau = 1/T_H\), and the on--shell condition can also be expressed as \(\partial_{r_h} F = 0\). Using the formalism outlined above, the new field and its corresponding unit vectors can be calculated
\refstepcounter{equation}
\begin{align}
\tilde{\phi}_r&=\partial_{r_h}F,\tag{\theequation a}\\\tilde{\phi}_{\Theta}&=-\mathrm { cot} \Theta \mathrm {csc} \Theta,\tag{\theequation b}
\end{align}
where \(\Theta\) satisfies \(0 \leq \Theta \leq \pi\). At \(\Theta = 0\) and \(\Theta = \pi\), the component \(\phi^\Theta\) diverges, with the direction of the vector pointing outward. By solving the equation \(\phi_r = \partial_{r_h} F\)=0, we can derive an equation in terms of \(\tau\). FIG. \ref{fig:PhaseF}(a) illustrates the zero points of the vector field \(\phi\) in the \(\tau-r_h\) plane. We observe three branches of black holes: the small and large black hole branches are stable, while the middle black hole branch is unstable. FIG. \ref{fig:PhaseF}(b) shows the unit vector field \(n\) at \(\tau = 10\). The zero points are located at \((0.454497, \pi/2)\), \((1.02267, \pi/2)\), and \((3.31018, \pi/2)\), respectively. We find that the topological charge is $+1$ for both the small and large black hole branches, whereas it is $-1$ for the middle black hole branch. Consequently, the total topological charge remains $+1$. Therefore, the
system has a similar topological classification to the charged RN-AdS black hole \cite{0029}.

\section{Conclusion}\label{Sec9}

In this study, we explored the effects of an antisymmetric Kalb--Ramond tensor field, which causes spontaneous Lorentz symmetry breaking, on the characteristics of a charged black hole in the presence of a cosmological constant. Our investigation centered on analyzing various properties, including the shadow radius, \textit{greybody} bounds, absorption and emission power, heat capacity, topological charge, and the optical features of the black hole. This research seeks to address a gap in existing literature and enhance our understanding of the consequences arising from this scenario of Lorentz symmetry breaking.

To begin with, we computed the shadow radius. In the specific scenario where \(M = 1\), we established the lower and upper limits for the parameter \(\bar{l}\) across three distinct values of \(Q\). Additionally, we examined characteristics that diverge from conventional solutions for charged black holes, such as the \textit{greybody} bounds and the associated absorption cross--section. For \(Q = 0.01\) and $\Lambda = -0.1$, we found that an increase in the parameter \(\bar{l}\) resulted in a decrease in the curves, while the emission power curve exhibited an upward shift as \(\bar{l}\) increased.

Importantly, we investigated the topological charge and the related topological phase transitions within this context. Through a detailed analysis of the metric, temperature, and free energy, we were able to determine the system's topological charge along with the associated phase transitions.

As a next step, we can extend our analysis to other configurations of black hole solutions that incorporate the Kalb--Ramond field, as discussed in Ref. \cite{19}. These concepts, along with additional ideas, are presently being explored and developed further.


\begin{acknowledgments}
This work is supported by the Doctoral Foundation of Zunyi Normal University of China (BS [2022] 07, QJJ-[2022]-314),  by the Long--Term Conceptual Development of a University of Hradec Králové for 2023, issued by the Ministry of Education, Youth, and Sports of the Czech Republic. Particularly,
A. A. Araújo Filho would like to thank Fundação de Apoio à Pesquisa do Estado da Paraíba (FAPESQ) and Conselho Nacional de Desenvolvimento Cientíıfico e Tecnológico (CNPq)  -- [150891/2023-7] for the financial support.
\end{acknowledgments}

\end{document}